# Channel Estimation for Massive MIMO Communication System Using Deep Neural Network

Zohreh Mohades ,Vahid TabaTaba Vakili

**Abstract**

In this paper we consider the problem of sparse signal recovery in Multiple Measurement Vectors (MMVs) case. Recently, ample researches have been conducted to solve this problem and diverse methods are proposed, one of which is deep neural network approach. Here, employing deep neural networks we have provided two new greedy algorithms in order to solve MMV problems. In the first algorithm, we create a stacked vector of measurement matrix columns and a new measurement matrix, which can be assumed as the Kronecker product of the primary compressive sampling matrix and a unitary matrix. Afterwards, in order to reconstruct sparse vectors corresponding to this new set of equations, a four-layer feed-forward neural network is applied. For the supervised learning, a set of learning data is required. The procedure of producing such data is also presented in the paper. In the second algorithm, this fact that sparse vectors have nonzero places in common, that is, vectors are jointly sparse is employed. Since recurrent neural networks, due to their feedback structure, are powerful tools for processing of sequence data, we apply them to extract joint sparsity structure of a sparse matrix containing joint sparse vectors. Indeed, in the latter algorithm, using recurrent neural networks a modified Subspace Pursuit (SP) algorithm, in which support signal is estimated and modified at each iteration, is presented. In order to compute the neural network parameters, we minimize the cost function of recurrent network utilizing Adam algorithm. In both suggested algorithms we learn networks just once and in the offline mode. Therefore, we do not worry about low speed of these algorithms, and the recovery speed of the proposed algorithms is the same as other conventional greedy algorithms. Comparing with the conventional reconstruction algorithms, the benefit of the proposed methods is shown. Finally, the proposed algorithms are utilized to estimate the pilot based channels of Massive MIMO systems.

Introduction

In recent years, sparse signal reconstruction has attracted much attention in the research community [1]. Specifically, consider the following compressive sensing (CS) model:

$$y = \mathbf{A}x \qquad (1)$$

where, $x$ is unknown sparse signal in $\mathbb{R}^N$, $\mathbf{A} \in \mathbb{R}^{M \times N}$ ($M \ll N$) is the sensing matrix and $y \in \mathbb{R}^M$ is the vector of measurements, where the goal is to reconstruct $x$ based on $y$ and $\mathbf{A}$. Since $N > M$, the system in (1) is an underdetermined system and has therefore infinitely many solutions. However, using the additional assumption that the signal $x$ is sparse, it is possible to uniquely and exactly recover $x$ via solving the $l_0$-miminization problem [1]:

$$\min \|z\|_0 \quad \text{subject to} \quad y = \mathbf{A}x \qquad (2)$$

Unfortunately, this combinatorial minimization problem is NP–hard in general [2]. To overcome this, many different algorithms have been proposed for finding the sparse solution. The most prominent approaches are divided in two different groups. The methods based on solving a convex optimization problem (like basis pursuit (BP) [3], Least Absolute Shrinkage and Selection Operator (LASSO) [4], and Basis Pursuit Denoising (BPDN) [3]) and also greedy iterative techniques such as orthogonal matching pursuit (OMP) [5], compressive sampling matching pursuit (CoSaMP) [6], and iterative hard thresholding (IHT) [7]. If the measurement matrix satisfies restricted isometry property (RIP) then one can apply reconstruction algorithms to recover compressed signal with acceptable accuracy. Since there is only one measurement vector, mentioned methods are appropriate in order to solve single measurement vector (SMV) problems in compressive sensing theory. In practice, we must recover a set of sparse vectors which usually are correlated. Therefore, we need to solve the problem of jointly reconstructing a set of sparse vectors.

Let $\mathbf{Y} = [y_1, \ldots, y_K] \in \mathbb{R}^{M \times K}$ be an observation matrix and $\mathbf{A} \in \mathbb{R}^{M \times N}$ be a sensing matrix so that there are some $\mathbf{X} \in \mathbb{R}^{N \times K}$ which $\mathbf{Y} = \mathbf{A}\mathbf{X}$. The problem of simultaneous sparse approximation is formulated as follows:

$$\min_{\mathbf{X}} \|\mathbf{X}\|_0 \quad s.t. \quad \mathbf{A}\mathbf{X} = \mathbf{Y} \qquad (3),$$

where, $\mathbf{X} = [x_1, \ldots, x_K]$ and $x_i$ is $i^{th}$ column of $\mathbf{X}$, $\|\mathbf{X}\|_0 = |\Omega|$, $\Omega = \text{supp}(\mathbf{X}) = \{1 \leq i \leq N | x^i \neq 0\}$ and $x^i$ demonstrates $i^{th}$ row of $\mathbf{X}$. Therefore, simultaneous sparse approximation problem means how to recover unknown matrix $X$, under the assumption that all columns of $X$ have joint support. Greedy algorithms [8] and Convex relaxation [9] are presented to solve such problems. Most of the algorithms proposed to solve MMV problems are based on, the rigorous assumption, joint sparsity model of the signals. In fact, this model rarely occurs in practice and signals only share a portion of complete joint sparsity model. From a more practically points of view, signals have not a pure sparse structure and therefore the sparsity based algorithms would lead to inexact results.

We, therefore, look for methods overcoming the aforementioned challenges. Deep neural networks (DNN) [10], as one of the most promising methods of machine learning, due to benefiting powerful learning from data, have been significantly considered in different areas like computer vision, speech recognition, image processing and sparse recovery. There are some literature addressing the usage of DNN in sparse recovery. In [11], in order to improve the reconstruction accuracy of structured signals a stacked denoising autoencoder (SDA), as an unsupervised feature learner, has been employed.

In [12], considering the multiple measurement vectors problem, in which the sparse signals are correlated, a convolutional deep stacking network (CDSN) has been proposed to extract information regarding the correlation of the sparse vectors and based on this information, a greedy algorithm has been presented to reconstruct the sparse signals from the measurement vectors. Instead using conventional CS reconstruction methods, either optimization based methods or greedy ones, Mousavi and et al have shown how to employ deep convolutional networks to reconstruct structured sparse signals, while improving both reconstruction accuracy and runtime [13]. In [14], a deep fully-connected network as a deep learning method has been utilized to reconstruct video frames from low dimensional samples. It has been shown how more accurate this method is relative to traditional CS reconstruction methods. In [15] and [16] two stacked de-noising autoencoders are introduced for reconstruction of Fetal electrocardiogram signals (FECG) and electroencephalogram (EEG) signals, respectively. In [15], signal recovery quality is improved by using fine tuning network via gradient descent-based back-propagation algorithm. In [17], an LSTM network is suggested to solve MMV problem and using experiment on real world data set, it is shown that the introduced procedure outperforms conventional sparse signal reconstruction methods.

Iterative sparse signal recovery algorithms, like OMP method, at each iteration, perform linear and nonlinear operations. Such iteration has been modeled with one layer of neural networks, and the iterations correspond to the depth of neural networks. This lets us use neural networks in the lieu of conventional sparse signal recovery algorithms and hope that in the presence of a set of large enough learning data achieve desirable results. Totally, from what presented in diverse literature, deep learning method, in comparison with conventional CS recovery methods, makes the signal recovery more accurate while, neglecting the learning phase, accelerates the reconstruction.

In this paper, to reconstruct the sparse matrix $\mathbf{X}$, each column of which is a sparse vector, from the measurement matrix $\mathbf{Y} = \mathbf{AX}$, a greedy algorithm, which uses deep neural networks to estimate and modify the support of the sparse matrix, is presented. In the reconstruction procedure, two famous neural networks, deep feed-forward neural network and recurrent neural network, are utilized. To state how fast the proposed algorithms might be, we should note that only the training phase, which is offline and done once, is severely time consuming and hence, the reconstruction algorithms would have the same running time as the conventional greedy algorithms do.

By comparing the proposed algorithms with the conventional sparse reconstruction methods, as expected, due to using deep neural networks, which are able to extract the data features, significant improvement on the exact sparse reconstruction is seen.

In Section II deep neural networks are briefly introduced. The proposed algorithms and the training data generation procedure are presented in the Section III. The proposed method is applied to the MaMIMO system in the Section IV. In Section V simulation results are illustrated. Section VI concludes the paper.

## II. Deep Neural Network

Deep neural networks [18], due to capability to extraordinarily extract data properties and model complex data, have been taken into consideration as powerful means in applications such as Machine Vision, Pattern Recognition and Image Processing. In this paper, two different structures of deep neural networks, i.e. deep feed-forward networks and recurrent neural networks are considered.

### II.1. Deep Feed-Forward Networks

In the feed-forward networks, information goes through layers directly, i.e. from input layer to the output layer without any feedback. A typical four layer feed-forward deep network is shown in the Fig. 1.

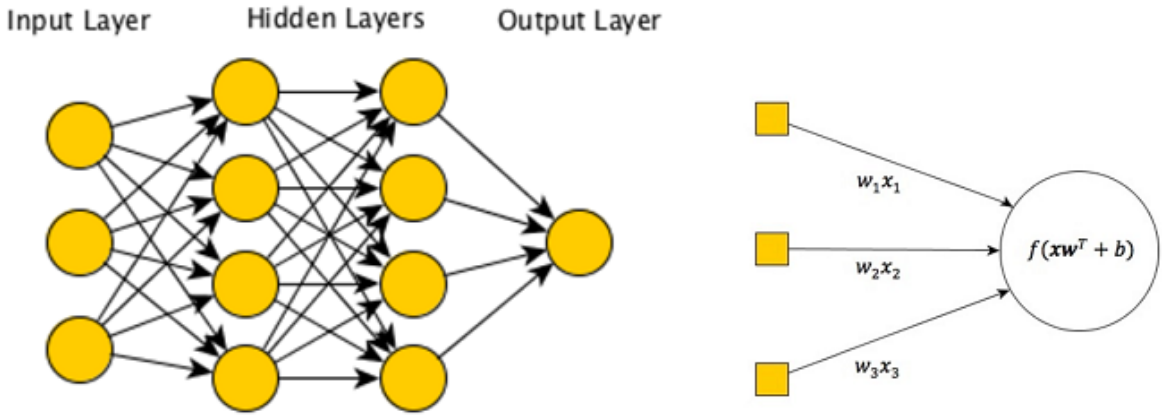

Fig.1. A typical four layer feed-forward deep network

As seen, each neuron of one layer receives its input data only from the neurons of the previous layers. Indeed, for the $j^{th}$ neuron at the $k^{th}$ layer, we have,

$$x_j^{k+1} = f\left(\sum_{i=0}^{n} x_i^k w_{i,j}^k\right) \quad (4)$$

where, $x_j^{k+1}$ is the $j^{th}$ neuron output at the $k^{th}$ layer, $x_i^k$ is the $i^{th}$ neuron output at the $(k-1)^{th}$ layer, is the connection weight of two neurons at the adjacent layers, and $f(.)$ is the network activation function.

### II. 2. Recurrent Neural Networks

Recurrent neural networks (RNN) are indeed kind of deep neural networks. This is easily seen when the RNN network is unfolded at time. Unlike the feed-forward networks, RNN networks own feedback connections making them capable to refer to the previous states, and hence, process different data sequences. A typical RNN network and the corresponding unfolded network have been illustrated in Fig. 2. The following equations are inferred from this figure.

$$o_t = \mathbf{W}_{IH} x_t + \mathbf{W}_{HH} h_{t-1} + b_h$$
$$h_t = f(o_t)$$
$$y_t = f(\mathbf{W}_{Ho} h_t + b_o) \quad (5)$$

where, $x_t$, $h_t$ and $y_t$ are the input, the hidden state and the output at time $t$, respectively and $h_{t-1}$ is the previous hidden state. Network weights are considered as $\theta = [\mathbf{W}_{IH}, \mathbf{W}_{HH}, \mathbf{W}_{Ho}, b_h, b_o]$.

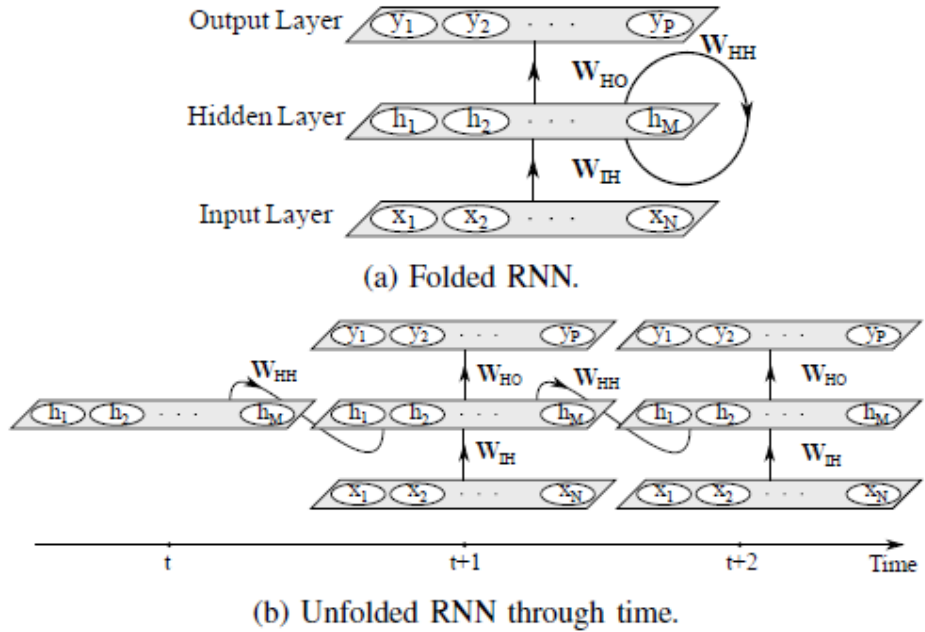

(a) Folded RNN.

(b) Unfolded RNN through time.

Fig.2. A typical RNN network and the corresponding unfolded network

### III. Proposed Methods

#### A. Modified BOMP Algorithm using deep feed-forward networks

In [19], BOMP has been proposed to reconstruct block-sparse signals. As mentioned, MMV problem deals with signals with joint sparse structure. Such a structure lets us rewrite the original problem in the form: $y = \Phi x$. where, $x = [x_1^T, x_2^T, ..., x_N^T]^T$ is a vector generated by deploying the rows of matrix X sequentially and also define matrix $\Phi = [\Phi[1], ..., \Phi[K]]$ as the Kronecker product of matrix $\mathbf{A}$ and the identity matrix $\mathbf{I}_{K \times K}$. This new vector benefits block sparsity structure, and hence, block sparse algorithms can be applied for.. Then, we consider the single sparse reconstruction problem from the equation $\mathbf{y} = \mathbf{\Phi}\mathbf{x}$, where $\mathbf{x} \in \square^{NK \times 1}$. To utilize neural network, the network is firstly trained at the offline state and the network parameters are adjusted. After that, applying an iterative algorithm sparse signal recovery is performed. Here, we use a 4-layer feed-forward neural network for each layer of which the activation function is considered to be tanh(.). In order to train, a training set including $T$ desired input and output pairs corresponding to $\{(r_i, x_i)\}_{i=1,\ldots,T}$ is defined. The way to generate the training data is discussed in the subsection A.1. To calculate the network parameters, i.e. $\mathbf{W}$ and b, the following cost function is minimized.

$$L(\mathbf{W}, b) = \frac{1}{T} \sum_{i=1}^{T} \left\| x_i - x_i^o \right\|_2^2 \qquad (6)$$

Where, $\mathbf{x}^o$ is the feed-forward neural network output obtained as follows.

$$a_1 = \tanh(\mathbf{W}_1 \mathbf{r} + \mathbf{b}_1), \quad a_2 = \tanh(\mathbf{W}_2 a_1 + \mathbf{b}_2), \quad a_3 = \tanh(\mathbf{W}_3 a_2 + \mathbf{b}_3), \quad x^o = \tanh(\mathbf{W}_4 a_3 + \mathbf{b}_4)$$

where, $\{\mathbf{W}_1 \in \square^{n_1 \times MK}, b_1 \in \square^{n_1 \times 1}\}$, $\{\mathbf{W}_2 \in \square^{n_2 \times n_1}, b_2 \in \square^{n_2 \times 1}\}$, $\{W_3 \in \square^{n_3 \times n_2}, b_3 \in \square^{n_3 \times 1}\}$, $\{\mathbf{W}_4 \in \square^{KN \times n_3}, b_4 \in \square^{KN \times 1}\}$ are the weights of the first, second, third and output layers, respectively, and $n_j$ is the number of nodes in $j^{th}$ layer.

To solve (6), propagation algorithm [20] is utilized. After obtaining the network parameters, the iterative algorithm in table I is applied to reconstruct the sparse signal x.

**Algorithm I**

**Input** $y_{MK \times 1}, \mathbf{\Phi}_{MK \times NK}$

**Output** $\hat{x}_{NK \times 1}$

**Initialization**

$i = 0, r^i = y$

**While** $\|r\| > \gamma$

1) Apply $r^i$ at deep feed-forward input to obtain $x^o$
2) $b^i =$ **find block with maximum norm in** $x^o$
3) $\arg\min_{\hat{x}^o[b^i]} \left\| y - \Phi[b^i]\hat{x}^o[b^i] \right\|_2$
4) $r^i = y - \Phi[b^i]x^o[b^i]$

$i = i + 1$

**End**

### A.1. Training Data Generation

In the case of block sparse signals, we generate the training data as follows. Firstly, measurement vector y is considered as the first input vector, and the corresponding output vector is a vector whose $b^{th}$ block is nonzero. This nonzero block is determined by finding the most correlated block with the vector y. After that, by calculating $\mathbf{r} = \mathbf{y} - \mathbf{\Phi}_i[b]x_i[b]$, the impact of this nonzero block is removed from measurement vector y and the remainder vector $r$ is obtained. Note that $\mathbf{\Phi}_i[b]$ is constructed by considering the columns of $\mathbf{\Phi}$ which corresponds to the $b^{th}$ block of vector x. Then, the remainder vector $r$ is considered as another input signal and the corresponding output vector is produced with the same manner of aforementioned output vector. This procedure is continued up to finding $k$ nonzero blocks. The set of obtained input and output vectors is the training data.

### B. Modified SP Algorithm Utilizing Deep Neural Networks

In [21] SP algorithm has been proposed to reconstruct sparse signals. In this paper we propose a modified SP algorithm using recurrent neural networks (MSPRNN) to reconstruct matrix $\mathbf{X}$ from (3). The algorithm is as follows. Firstly, $k$ greatest elements of absolute value of each column of $\mathbf{A}^*\mathbf{Y}$ are considered as initial support ($T_0$) corresponding to each column of matrix $\mathbf{X}$. Considering this support, the residual vector for each column of X is obtained as follows,

$$\mathbf{R}_0(i) = \mathbf{Y}(:,i) - \mathbf{A}_{T_0(i)} \left( \mathbf{A}^*_{T_0(i)} \mathbf{A}_{T_0(i)} \right)^{-1} \mathbf{A}^*_{T_0(i)} \mathbf{Y}(:,i) \quad (7)$$

where $i$ is the column index. These residual vectors are considered as recurrent neural network input. The support is updated in the step (7), which is the union of the previous support and the $k$ indices of the neural network output. After that, in step (8), matrix $\mathbf{X}$ is estimated. The support vector is modified in step (9). This algorithm is repeated till the desired criterion, i.e. residual norm gets less than a given threshold, is satisfied.

**Algorithm II**

**Input** $\mathbf{Y}_{M \times K}, \mathbf{A}_{M \times N}, k$

**Output** $\hat{T}, \hat{\mathbf{X}}_{N \times K}$

**Initialization**

$\hat{\mathbf{X}} = 0$

**For** $i = 1:K$

$$\hat{T}_0(i) = \{k \text{ greatest index of } \mathbf{A}^*\mathbf{Y} \text{ in every vectors}\}$$

$$\mathbf{R}_0(i) = \mathbf{Y}(:,i) - \mathbf{A}_{T_0(i)}\left(\mathbf{A}^*_{T_0(i)}\mathbf{A}_{T_0(i)}\right)^{-1}\mathbf{A}^*_{T_0(i)}\mathbf{Y}(:,i)$$

**End**

$i = 0$

**While** $\|\mathbf{R}_i\|_F > \gamma$

**For** $j = 1:K$

    5) **Apply** $\mathbf{R}_i(j)$ at RNN input to obtain $\mathbf{V}_i(j)$

    6) **Index**= $k$ index of $Max\{|\mathbf{V}_i(j)|\}$

    7) $\hat{T}_i(j) = \hat{T}_{i-1}(j) \cup \text{Index}$

    8) $\hat{\mathbf{X}}_{\hat{T}_i}(:,j) = \mathbf{A}_{\hat{T}_i}^\dagger \mathbf{Y}(:,j)$

    9) $\hat{T}_i(j) = \{k \text{ greatest index of } \hat{\mathbf{X}}(:,j)\}$

    10) $\mathbf{R}_i(:,j) = \mathbf{Y}(:,j) - \mathbf{A}_{\hat{T}_i(j)}\hat{\mathbf{X}}_{\hat{T}_i(j)}(:,j)$

**End**

$i = i + 1$

**End**

## B.1. Training Data Generation

In order to generate the training data, i.e. input and target, where in the sparse signal recovery problem input and target stand for residual vectors and the corresponding sparse signal vectors, respectively, we act as follows. For each column of matrix $\mathbf{X}$, say $i^{th}$ column, a vector $\mathbf{x} \in \square^N$ with elements equal to 1, corresponding to the $k$ largest absolute values of the $i^{th}$ column of $\mathbf{A}^*\mathbf{Y}$, and, zero elements at the other positions, is generated. Such vector is the input and the corresponding vector from $\mathbf{Y} = \mathbf{A}\mathbf{X}$ is the target vector. Then, we adopt the same procedure but just replacing the residual, which is obtained as $\mathbf{R} = \mathbf{Y} - \mathbf{A}_\mathbf{x}\left(\mathbf{A}^*_\mathbf{x}\mathbf{A}_\mathbf{x}\right)^{-1}\mathbf{A}^*_\mathbf{x}\mathbf{Y}$, instead $\mathbf{Y}$. This procedure is continued till the residual norm reduces as much as desired.

## VI. CHANNEL ESTIMATION IN MASSIVE MIMO WITH PROPOSED ALGORITHMS

In this section we use our proposed framework for pilot-based channel estimation problem in downlink frequency division duplex (FDD) massive MIMO systems [22]. To achieve various performance gains in massive MIMO systems the knowledge of the channel plays an important role in these systems. Pilot-based channel estimation techniques require the transmission of pilots that are known to the receiver. With the use of these pilots it is possible to extract the channel information at the receiver. In this paper we consider a system consisting of one BS with M transmitting antennas and one user with N receive antennas $(M \gg N)$. In the training phase, BS sends T training symbols on its M antennas. The received signal at the user is

$$\mathbf{Y} = \mathbf{HS} + \mathbf{N} \tag{8}$$

Where, $\mathbf{S} \in \mathbb{C}^{M \times T}$ is the pilot matrix as $tr(\mathbf{S} \times \mathbf{S}) = P\mathrm{T}$, where $P$ denotes the transmitted power from the BS, $\mathbf{H} \in \mathbb{C}^{N \times M}$ is the channel matrix and $\mathbf{N} \in \mathbb{C}^{N \times T}$ is the additive complex Gaussian noise with zero mean and unit variance. The channel matrix $\mathbf{H}$ in the angular-domain can be represented as

$$\mathbf{H} = \mathbf{A}_R \mathbf{H}^a \mathbf{A}_T^H \tag{9}$$

where, $\mathbf{A}_R \in \mathbb{C}^{N \times N}$ and $\mathbf{A}_T \in \mathbb{C}^{M \times M}$ denote the angular domain transformation unitary matrices at the transmit and receive sides, respectively, and $\mathbf{H}^a \in \mathbb{C}^{N \times M}$ is the angular domain channel matrix. Due to BS is usually elevated high with limited scatterers around and users are located at low elevation with relatively rich number of local scatterers, matrix $\mathbf{H}^a$ have two properties:

**Property 1:** $\mathbf{H}^a$ is sparse.

**Property 2:** All of the columns $\mathbf{H}^a$ have the same support.

From the mentioned properties, we can apply our proposed algorithms for the reconstruction of the channel matrix. First, the signal model (8) can be re-written as

$$\begin{aligned} \mathbf{Y} &= \mathbf{A}_R \mathbf{H}^a \mathbf{A}_T^H \mathbf{S} + \mathbf{N} \\ \mathbf{Y}^H \mathbf{A}_R &= \mathbf{S}^H \mathbf{A}_T \mathbf{H}_a^H + \mathbf{N}^H \mathbf{A}_R \\ \bar{\mathbf{Y}} &= \bar{\mathbf{A}} \bar{\mathbf{X}} + \bar{\mathbf{N}} \end{aligned} \tag{10}$$

where

$$\bar{\mathbf{Y}} = \mathbf{Y}^H \mathbf{A}_R, \quad \bar{\mathbf{A}} = \mathbf{S}^H \mathbf{A}_T, \quad \bar{\mathbf{X}} = \mathbf{H}_a^H, \quad \bar{\mathbf{N}} = \mathbf{N}^H \mathbf{A}_R \tag{11}$$

Now the CSIT estimation is equivalent to standard CS recovery problem and we apply Algorithm I and Algorithm II for solving equation (10).

**V. SIMULATION RESULTS**

In this section we evaluate the performance of the proposed methods. We compare our results with some well-known traditional MMV reconstruction methods, i.e. Simultaneous Orthogonal Matching Pursuit (SOMP), Group LASSO (G-LASSO), and Subspace Pursuit (SP), in sense of normalized mean square error (NMSE) defined in the following.

$$NMSE = \frac{\|\hat{\mathbf{H}} - \mathbf{H}\|}{\|\mathbf{H}\|},$$

where, $\mathbf{H}$ is the original channel matrix and $\hat{\mathbf{H}}$ is the reconstructed one, and the norm is Frobenius norm. Each of the traditional MMV reconstruction methods owns properties encouraging us to use them. SOMP method inherits the properties of OMP method which is one of the fastest traditional reconstruction methods. G-LASSO method benefits an extra-norm term relative to the LASSO method which is useful to consider the joint sparsity structure of the signals. The last method, SP, is one of the most promising compressive sensing construction methods which efficiently extracts the subspace of the signal. In the Massive MIMO channel estimation case in which joint sparsity is the attribute of the signals, this method would work more reliable to find the subspaces of the signals. For example, suppose that one signal suffers from low SNR, and consequently, the probability of finding the true support decreases significantly; in this case it is possible to extract the exact support of such a signal by imposing the same support distribution with the other signals whose SNR is adequate enough to estimate their supports. In fact, we have utilized SP method to reconstruct signals with high sparsity orders so that the signals could be barely reconstructed unless the same support distribution constraint is applied. This, somehow, can be considered as a generalization of the G-LASSO method.

Now, we talk about the situations that the sparse signal reconstruction is investigated. We consider a narrow band (flat fading) point-to-point massive MIMO system with one BS and one user, where the BS and user have M = 144 and N = 4 antennas, respectively, and the length of the training pilot is T=72. We use the 3GPP spatial channel model (SCM) [23] to generate the channel coefficients and we consider that the user has a rich local scattering environment [24]. As proposed, we consider two different neural networks, i.e. recurrent neural network and feed-forwad neural network. In order to train RNN and feed-forward networks we use 12000 and 15000 pairs of generated training data respectively, as discussed in Sections (III.A.1 and III.B.1). For training feed-forward network we generated 15000 training sample pairs (input, target) where an input vector of length 288 and a target vector of length 576 with at most 72 non-zero elements with value 1. We used a standard four-layers fully connected feed-forward network with $n_h = 256$. In training of RNN, each pair consists of an input vector of length 72 and a target vector of length 144 with at most 18 non-zero elements with value 1. For RNN, during training, we used one layer, with $n_h = 1024$ and 30 iterations of parameter updates and 12000 pairs (input, target) including 4 different parts, corresponding to the user's four antennas, employed sequentially. In other words, first 3000 pairs contain the training data corresponding to the first antenna of the user and similar correspondence is considered for the other pairs. As stated, the training data is generated using SP method which may converge with different number of iterations for different signals and this difference would impose a difficulty with simulations. Because, we use a tensor format, i.e. input data with size of $72\times4\times12000$ and target data with size of $144\times4\times12000$, in our simulations and we need to have the same number of pairs per each part (as pointed out we have 4 parts corresponding to each antenna of a given user). To avoid the different number of pairs per part, we removed some of the latest pairs of each part to unify the number of pairs at each pair. The pilot utilized to estimate the channel is generated randomly with standard normal distribution.

The results are depicted in Figs. 1, 2, 3 & 4, where, Figs. 1 and 2 show the normalized mean square error (NMSE) achieved by the two methods for different SNRs. As seen, our proposed algorithms significantly outweigh all the other mentioned ones except the SP method which is slightly outperformed. The reason we can conceive about such better performance is that, in our methods, the neural network not only has extracted the sparsity model of the signals but also the structure of the sparsity, where, virtually all the nonzero elements are alongside each other. In Figures 2 and 4, we compare the NMSE of the estimated channel versus the length of the training pilot T, under transmit power P = 35dB. As observed from Figs. 3 and 4, using proposed methods improves the reconstruction performance compared to other methods discussed in this paper. Totally, by comparing the proposed method with some well-known conventional reconstruction methods, it is seen there is a great desire to adopt such approach to reconstruct the sparse signals in the case of Sparse Channel Estimation.

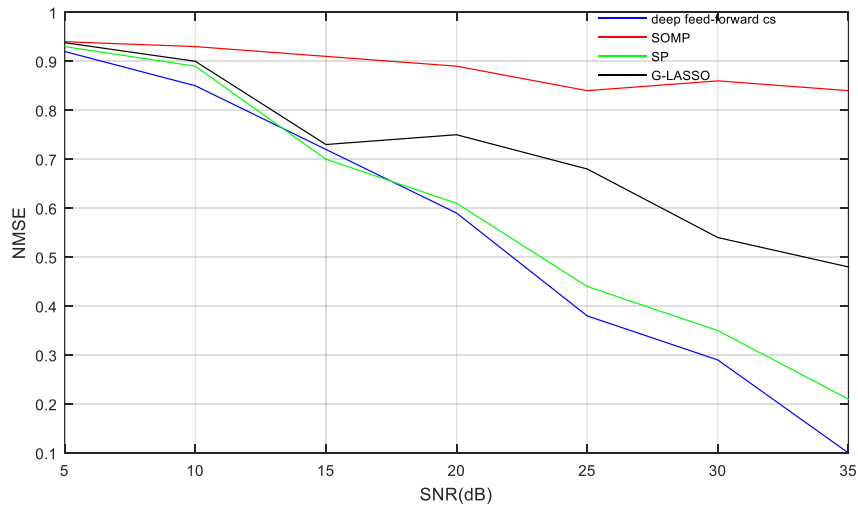

Fig.3. NMSE of the different method versus SNR (Comparison with Algorithm I)

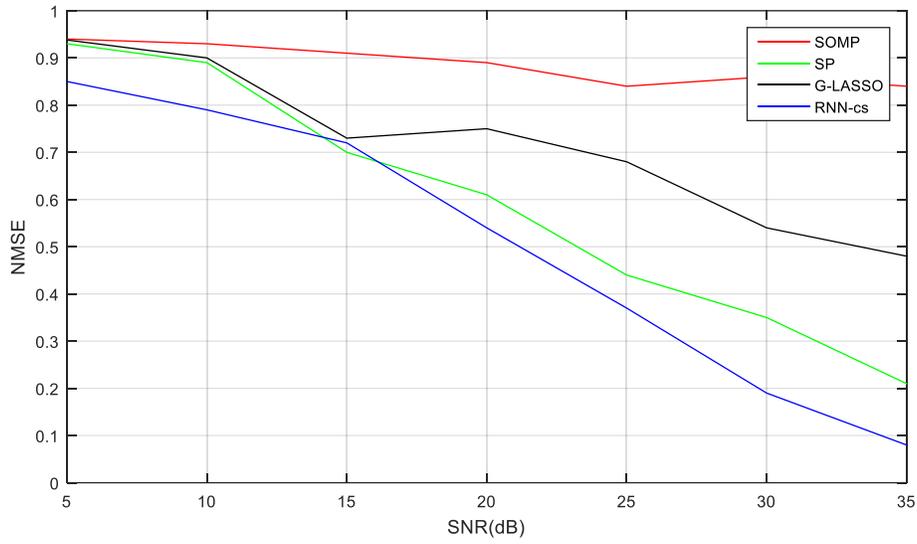

Fig.5. NMSE of the different method versus SNR (Comparison with Algorithm II)

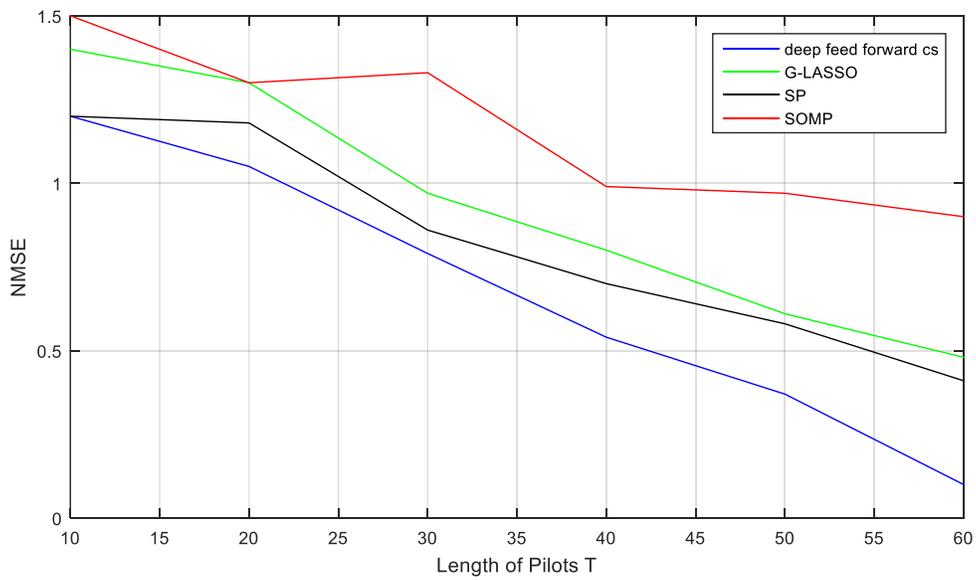

Fig.4. NMSE of the different method versus the pilot training length T (Comparison with AlgorithmI)

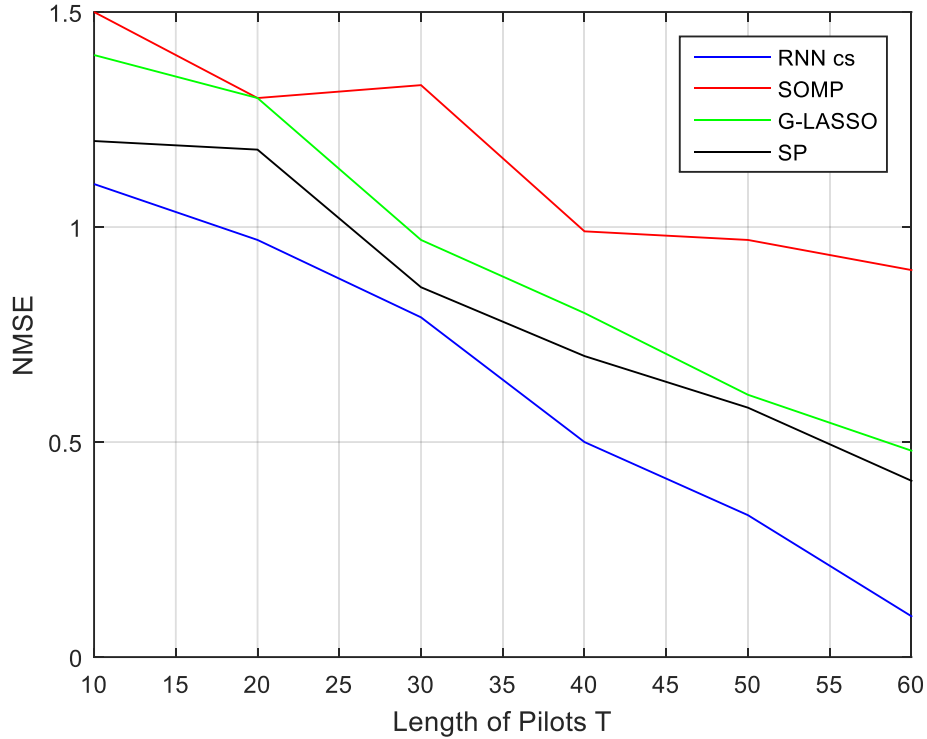

Fig.6. NMSE of the different method versus the pilot training length T (Comparison with AlgorithmII)

## VI. CONCLUSIONS

In this paper we presented two algorithms to reconstruct sparse vectors for the MMV problems. The proposed methods, by using deep neural network, learn the structure of sparse vectors and then, applying such learned networks, reconstruct the original sparse signals. We applied the proposed methods to channel estimation problem in massive MIMO systems and showed that the proposed methods outperform the conventional MMV methods.

Some future works can be more profoundly investigated include, first, considering prior information to the signal and evaluating the performance of the proposed algorithms under such case, and second, finding the phase transition diagrams which exhibit the number of required measurements to exactly reconstruct sparse signals with different number of sparsity.